## Statement of Contributions

The first contribution of this research is capturing driver behavior in congested and incident-prone situations, thus incorporating drivers' risk-taking attitude in the model equations.    The model formulated in this paper does not exogenously impose safety constraints to prevent accidents from occurring. Models used in practice typically preclude accidents, contrary to real-life situations. One more implication of this contribution is capturing that drivers do not perfectly register existing stimuli without subjectively weighing different alternatives based on their personality (aggressive versus conservative drivers). This allows risky behavior as an inherent result of the model. Moreover, the corresponding acceleration choice emerges as a probabilistic decision making process facing uncertainty; the method by which the resulting accident causing behavior is weighed can be calibrated based on recorded traffic data.

From a practitioner stand point, the main challenge in realizing the above contribution and incorporating the corresponding parameters is the degree of complexity that would be added to the eventual model which would preclude its usefulness in actual practice. Accordingly, the second contribution of this research is to put forward a "logic" that is robust enough to advance the state of knowledge related to the driving task but simpleand fastenough so that it can be readily implemented, calibrated and validated. The resulting model is intended to provide a competitive stochastic alternative to existing simpler models that lack cognitive dimensions.

# Highlights

- A car-following model is formulated using a prospect theory value function.

- Risk taking and proneness to accidents are quantified through the offered framework.

- Realistic traffic properties are reproduced after extensive numerical approximations.

- While capturing heterogeneity, the model is calibrated against NGSIM trajectory data.

- Traffic break-down followed by congestion flow-density data scattering are observed

# From Behavioral Psychology to Acceleration Modeling: Calibration, Validation, and Exploration of Drivers' Cognitive and Safety Parameters in a Risk-Taking Environment

**Samer H. Hamdar***
hamdar@gwu.edu
Department of Civil and Environmental Engineering
**The George Washington University**
801 22nd Street, NW
Academic Center #631
Washington, DC 20052
Phone: (202) 994-6652
Fax: (202) 994-0127

**Hani S. Mahmassani**
masmah@northwestern.edu
Department of Civil and Environmental Engineering
**Northwestern University**
The Transportation Center
Chambers Hall, 600 Foster Street
Evanston, IL 60208
Phone: (847) 491-5017
Fax: (847) 491-3090

**Martin Treiber**
treiber@vwi.tu-dresden.de
Institute for Transport and Economics
**Technische Universität Dresden**
Andreas-Schubert-Straße 23
D-01062, Germany
Phone: +49-351-463-36794
Fax: +49-351-463-36809

December 2013

# From Behavioral Psychology to Acceleration Modeling: Calibration, Validation, and Exploration of Drivers' Cognitive and Safety Parameters in a Risk-Taking Environment

Samer H. Hamdar* (hamdar@gwu.edu) , Hani S. Mahmassani and Martin Treiber

**ABSTRACT**

We investigate a utility-based approach for driver car-following behavioral modeling while analyzing different aspects of the model characteristics especially in terms of capturing different fundamental diagram regions and safety proxy indices. The adopted model came from an elementary thought where drivers associate subjective utilities for accelerations (i.e. gain in travel times) and subjective dis-utilities for decelerations (i.e. loss in travel time) with a perceived probability of being involved in rear-end collision crashes. Following the testing of the model general structure, the authors translate the corresponding behavioral psychology theory - prospect theory - into an efficientmicroscopic traffic modeling with more elaborate stochastic characteristics considered in a risk-taking environment.

After model formulation, we explore different model disaggregate and aggregate characteristics making sure that fidelity is kept in terms of equilibrium properties. Significant effort is then dedicated to calibrating and validating the model using microscopic trajectory data. A modified genetic algorithm is adopted for this purpose while focusing on capturing inter-driver heterogeneity for each of the parameters. Using the calibration exercise as a starting point, simulation sensitivity analysis is performed to reproduce different fundamental diagram regions and to explore rear-end collisions related properties. In terms of fundamental diagram regions, the model in hand is able to capture traffic breakdowns and different instabilities in the congested region represented by flow-density data points scattering. In terms of incident related measures, the effect of heterogeneity in both psychological factors and execution/perception errors on the accidents number and their distribution is studied. Through sensitivity analysis, correlations between the crash-penalty, the negative coefficient associated with losses in speed, the positive coefficient associated with gains in speed, the driver's uncertainty, the anticipation time and the reaction time are retrieved. The formulated model offers a better understanding of drivers behavior, particularly under extreme/incident conditions.

## 1. INTRODUCTION

In 2008, an acceleration-based car-following model was proposed that incorporates the risk-taking attitudes of drivers and uses prospect theory to evaluate the perceived consequences of applying different acceleration rates, a probability of collision and a crash penalty term are explicitly introduced in the formulation (Hamdar et al., 2008). This paper builds on this approach for exploring the characteristics of the formulated car-following model in terms of its ability to capture congestion regions, equilibrium characteristics, inter-driver heterogeneity and collective accident-prone behaviors on a freeway section. Being calibrated against real-life trajectory data (FHWA – 2004 a,b,c), different bottleneck and incident scenarios are modeled: bottleneck scenarios are tested via deceleration exerted by the leader and on-ramp merging; incident scenarios are tested via rear-end collision and fixed object crashes. Special interest is given to studying the resulting fundamental diagram especially traffic breakdown and the congestion disturbances. The effect of both psychological factors and execution/perception errors on the accidents number and their distribution along a freeway length is also studied. Through sensitivity analysis, insights into the relationships between the crash-penalty, the negative coefficient associated with losses in speed, the positive coefficient associated with gains in speed, the driver's uncertainty, the anticipation time and the reaction time are provided.

Theobjective of this research is to offer a comprehensive study of how the model performs in describing homogeneous and heterogeneous traffic flow under different traffic conditions (including extreme/incident conditions) giving a better insight into the psychological/cognitive reasoning adopted. Accidents are created as an inherent result of the utility function by relaxing some of the usually adopted safety constraints.

The first contribution of this research is capturing driver behavior in congested and incident-prone situations, thus requiring incorporating drivers' risk-taking attitude in the model equations. The model formulated in this paper does not exogenously impose safety constraints to prevent accidents. Models used in practice typically preclude accidents, contrary to real-life situations. One more implication of this  contribution is capturing that drivers do not perfectly register existing stimuli without subjectively weighing different alternatives based on their personality (aggressive versus conservative drivers). This allows risky behavior as an inherent result of the model. Moreover, the corresponding acceleration choice emerges as a probabilistic decision making process facing uncertainty; the method by which the resulting accident causing behavior is weighed can be calibrated based on recorded traffic data.

From a practitioner stand point, the main challenge in realizing the above contribution and incorporating the corresponding parameters is the degree of complexity that would be added to the eventual model which would preclude its usefulness in actual practice. Accordingly, the second contribution of this research is to put forward a "logic" that is robust enough to advance the state of knowledge related to the driving task but simpleand fastenough so that it can be readily implemented, calibrated and validated. The resulting model is intended to provide a competitive stochastic alternative to existing simpler models that lack cognitive dimensions.

In other words, the main challenge faced is translating the behavioral psychology prospect theory into a concise acceleration formulation given the importance of such structure for the calibration and the simulation exercise; this challenge is faced through the use of a GA calibration heuristic that allows calibrating the model for each "feasible" vehicle and attempting to capture a heterogeneity pattern. The structure of this paper will then follow; a background review on incidents and pertinent car-following models is presented in the following section. The framework of the work is shown in the third section where the corresponding car-following model is presented. The review and the framework will motivate testing the model in terms of equilibrium conditions. After calibrating the model, the fifth section includes the simulation results and the corresponding data analysis before concluding with some future research needs.

## 2. BACKGROUND REVIEW

In the year 2000, the monetary cost related to traffic accidents reached 230.6 billion USD (U. S. Dollars) in the U.S.A., only (NHTSA, 2007). Based on the National Highway Traffic Safety Agency (NHSTA) studies, 5 accident types of interest can be identified: 1) rear impacts (29.6% of US accidents), 2) angle or side impacts (28.6 % of US accidents), 3) fixed object crashes (16.1 % of US accidents), rollovers (2.3% of US accidents), head-on collisions (2 % of US accidents) and collision with pedestrians/bicyclists (1. 8 % of US accidents) (3). In car-following, the focus is on the tailgating behavior that may lead to rear-end collisions (Type 1). However, existing car-following models are designed to be accident free and therefore are, by definition, unsuited to capture driver behavior during incident scenarios (FHWA, 2004; Hamdar and Mahmassani, 2008); Moreover, a limited amount of research focused on the cognitive and risk-taking attitudes in driver behavior including the heterogeneity aspect that leads to scattering of flow-density data points and a more favorable environment for incidents.

The main assumption in "standard" car-following models is that the behavior of the following vehicle (e.g. change in acceleration) is directly related to a stimulus observed/perceived by the driver, defined relative to the lead vehicle (e.g. difference in speeds, headways etc.). This idea was adopted in the car-following models of Chandler, Gazis and Herman (Chandler et al., 1958, Gaziz et al., 2959 and Herman et al., 1959), known as the General Motor (GM) models. These first models are not complete in the sense that they are not applicable to all traffic situations including, e.g., free traffic or approaching standing vehicles or obstacles. Later investigations proposed improved models by introducing a "safe" time headway and a desired speed. The Gipps model (Gipps, 1981), and the intelligent-driver model (IDM) (Treiber et al., 2000) contain intuitive parameters that can be related to the driving style such as desired accelerations, comfortable decelerations, and a desired "safe" time gap. Furthermore, they include braking strategies that prevent accidents under a given heuristic. Subsequent studies have extended these models, by introducing additional parameters intended to capture dimensions such as anticipation, learning, and response to several vehicles ahead. Other models such as the human driver model (HDM) (Treiber et al., 2006) also model human deficiencies, including variable reaction times and the size and persistency of estimation errors of the input stimuli depending on the traffic situation. The Wiedemann model captures the indifference of the drivers to small changes in the stimuli. It also allows different execution modes including emergency braking (Wiedemann and Reiter, 1992).

Calibrating the above models need different levels of effort based on data availability, the number of parameters to calibrate, the calibration method and the model structure. For example, calibrating the Wiedemann Model requires estimating 18 parameters found in 17 different equations. On the other hand, in the IDM model, drivers behavior is captured by one equation with 5 parameters to estimate.

Before recent developments in collecting microscopic data (allowing the NGSIM research effort, FHWA 2004, 2005 a, b and c), a rare amount of data was available to calibrate the existing microscopic car-following models stated above. One of these data sets was collected by wire-linked vehicles on a test track at the General Motors Technical Center (Hamdar et al., 2009). Another technique was by using a camera attached to a helicopter. The gathered pictures were input to a time consuming manual processing system (Ossen and Hoogendoorn, 2007). Lately, image processing software and Differential Global Positions System (DGPS) have become available. This gave new tools to researcher to collect more accurate and detailed individual driver information.

Once the data is available, different calibrations techniques can be applied. In the "traditional" model calibration process, the "car-following" model parameters need to be adjusted until an acceptable (qualitative and quantitative) match is found between the simulated model dynamics and the observed drivers' behavior. Engineering judgment and trial-and-error methods are still widely used especially in the industry (Chu et al., 2004). More systematic approaches including the gradient method (Hourdakis et al., 2002) and Genetic Algorithm (Cheu et al., 1998) address the model calibration procedure as an optimization problem: a combination of parameter values are searched so an objective function (error term) is minimized. Lately, most research is oriented to capture intra and inter driver heterogeneity and time correlation in the parameters estimates (Ossen and Hoogendoorn, 2007).

## 3. FRAMEWORK: the Car-Following Model

In this section, the general structure of the stochastic acceleration model is introduced. The detailed implementation details and the individual parametric equations of this model can be seen in (Hamdar et al., 2008). Some analyticaland numericalderivations are not presented in this paper for conciseness.

In the free-flow regime, the main factor governing the acceleration behavior of adriver is his or her desired speed $v_0$ (Gipps, 1981). The acceleration applied by a driver toreach this speed starting at a speed $v$ and having a maximum possible acceleration value $a_{max}$ is given by

$$\dot{v}_{free} = a_{max}\left(1 - \frac{v}{v_0}\right). \qquad \textit{Equation 1}$$

In other words, the acceleration is always to be restricted by a free-flow acceleration function where $\frac{dv}{dt} = \dot{v}_{free} + \dot{v}_{int}$ and $\dot{v}_{int}$ is the acceleration adopted when interactions between vehicles is present.

The acceleration in dense or congested traffic is mainly controlled by interactions with the leading vehicle. In this model, the acceleration is modeled by a stochastic process that is characterized by the following:

1- The expected acceleration value $\langle\hat{a}\rangle(t)$

2- the variance $\sigma_a^2(t)$

3- and by the correlation time $\tau_{corr}$

At each given time, the stochastic acceleration $\dot{v}_{int}$ is distributed according to a continuous logit model whose distribution function is conditioned to the actual speed $v(t)$, the space gap $s(t)$ to the leader, and the relative speed to the leader $\Delta v(t)$ to the leader ($\Delta v > 0$ when approaching): $\dot{v}_{int} \sim \text{LOGIT}(s, v, \Delta v)$.

The conditional probability density $f_{int}(a \mid s, v, \Delta v)$ of the Logit model is given by

$$f_{int}(a \mid s, v, \Delta v) = \frac{e^{\beta U(a; s, v, \Delta v)}}{\int e^{\beta U(a'; s, v, \Delta v)} da'}. \qquad \textit{Equation 2}$$

The (generalized) utility $U$ of the model is composed of the generalized (or perceived) prospect-theoretic acceleration utility $U_{PT}(a)$ whose form is derived by the prospect theory, and a penalty $U_{crash}(a)$ for the risk of accidents:

$$U(a; s, v, \Delta v) = U_{PT}(a) + U_{crash}(a; s, v, \Delta v). \qquad \textit{Equation 3}$$

We specify the utility component by

$$U_{PT}(x) = x * \left(w_m + 0.5 * (1 - w_m) * (\tanh(x) + 1)\right) * \left(1 + x^2\right)^{0.5*(\gamma-1)}, \qquad \textit{Equation 3.1}$$

$$\textit{where } x = \frac{a}{a_0},$$

and

$$U_{crash}(a; s, v, \Delta v) = p_c w_c\, k(v, \Delta v).\qquad \textit{Equation 3.2}$$

In the prospect theoretic utility $U_{PT}$, a weighing function is adopted to evaluate the subjective utilities of different accelerations (Tversky and Kahnemann, 1986). The gains and losses are expressed as a function of acceleration, or, equivalently, in terms of expected speed gains and losses over a specific period of time (Figure 1). The non-varied model parameter $a_0$ indicates the subjective scale of the acceleration: accelerations $|\dot{v}_{int}| < a_0$ are considered to be "near the reference point" leading to increased sensitivity (Figure 1). Otherparameters of interest in the corresponding value function are the weight associated with negative acceleration ( $w^-$ or $w_m$ ) and the nonlinear sensitivity component $\gamma$. The weight associated with the gains ( $w^+$ ) is assumed to be one, so is the weighting of losses relative to that of gains(relative measure between $w^+$ and $w_m$ ).In other words, for $\gamma$ = 1 (no increased sensitivity at the reference acceleration), the function expressed in Equation 3.1 has linear asymptotes just retaining the different positive and negative weighing, and a smooth transition of width $a_0$around zero. Additionally, when $w_m$= 1, the utility becomes linear as $U_{PT}(x) = x = \dfrac{a}{a_0}$.

The second term on the right hand side of Equation 3 denotes the crash-related utility. In contrast to $U_{PT}(a)$ which is monotonously increasing with the acceleration $a$, $U_{crash}$ is monotonously decreasing with $a$ since a higher acceleration, and the ensuing higher future speed, increases the risk of rear-end collisions. The utility $U_{crash}$ consists of the estimated probability of a crash $p_c$, a seriousness term reflecting the expected adverse consequences of an accident$k(v, \Delta v)$ and a crash weight $w_c$. The gradient of the crash utility is given explicitly by Equation 13.The estimated crash probability $p_c$ is the probability of a rear-end collision within the time horizon $\tau$ assuming that i) the chosen acceleration $a$ of the follower remains unchanged within this interval, ii) the speed of the leader is constant, and iii) this speed is only known imprecisely in terms of an unbiased Gaussian distribution of relative error (variation coefficient) $\alpha$.In other words, when estimating the crash probability, drivers are assumed to predict the future position of a leader where the variation of this speed is dictated by an estimation uncertainty $\sigma(v_l) = \alpha v_l$ of the speed of the leader$v_l$.On the other hand,$k(v, \Delta v)$increases both with the speed $v$ and the approaching rate $\Delta v$ at present. For simplicity, we neglect these variations and we set the term $k(v, \Delta v)$ to be equal to a constant weight.Regarding$w_c$, a higher $w_c$corresponds to conservative individuals while a lower corresponds to drivers willing to take a higher risk.

Notice that we assume the utility to be dimensionless. Furthermore, its derivative $U' \equiv \dfrac{dU}{da}$ with respect to acceleration is of the order of $1/a_0$ where $a_0 = 1\text{m/s}^2$. Consequently, $\dfrac{1}{\beta}$ has the order of magnitude of the intra-driver uncertainties of the acceleration.

### 3.1. Expectation Value

For sufficiently high values of $\beta$ (which we will assume henceforth), the expectation value $\langle \hat{a} \rangle$ of the distribution for $\dot{v}_{int}$ can be approximated by its mode value, i.e., by the acceleration at the maximum of its probability density:

$$\langle \hat{a} \rangle (s, v, \Delta v) = \int a f_{int}(a \mid s, v, \Delta v) da \approx a*(s, v, \Delta v) = \arg \max U(a; s, v, \Delta v) . \quad \textit{Equation 4}$$

As usual, the value $a*(t)$ for maximum utility can be determined by the condition:

$$U'(a*; s(t), v(t), \Delta v(t)) = 0 . \qquad \textit{Equation 5}$$

The dependencies between $U_{PT}(a)$ and $U_{crash}(a; s, v, \Delta v)$ usually lead to a unique maximum of the generalized utility at some acceleration $a*$. The condition in Equation 5, however, generally will be satisfied for two values of the acceleration, where the higher one pertains to a minimum of the utility (unsafe driving mode). Therefore, as mentioned and explained in (Hamdar et al., 2008), a good initial guess for $a*$ is essential.

### 3.2. Variance and Correlation Time

Assuming again a sufficiently large value of the Logit uncertainty parameter $\beta$, the variance of the distribution characterized by the probability density in Equation 2 can be calculated with the method of the asymptotic expansion. The result is:

$$\sigma_a^2(t) = \frac{-1}{\beta U''(a*(t), s(t), v(t), \Delta v(t))}, \qquad \textit{Equation 6}$$

where the second derivative $U''(a*, s, v, \Delta v)$ can be calculated analytically. The correlation time is given directly by the model parameter $\tau_{corr}$.

The total number of parameters that need to be calibrated is seven. These parameters are presented in Table 1. The corresponding prospect theoretic utility function $U_{PT}$ may be observed below:

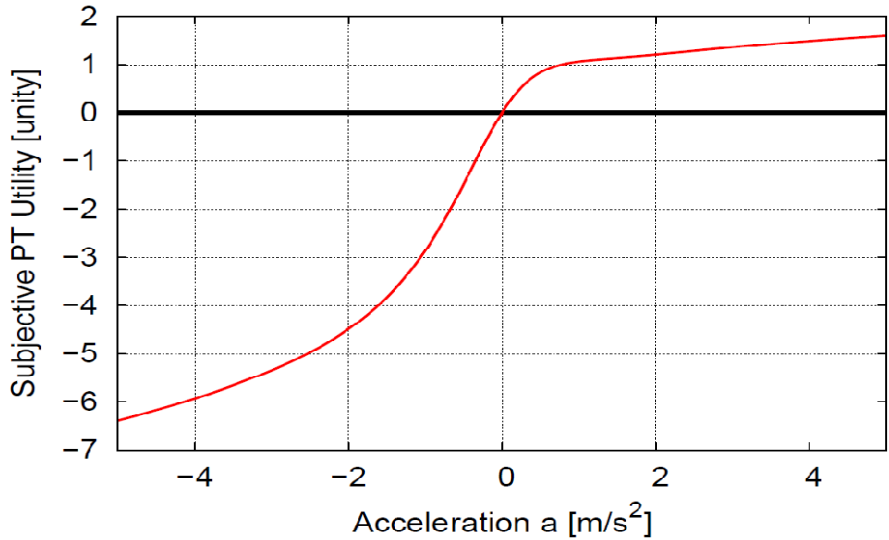

**Figure 1: Prospect theoretic utility function $U_{PT}$ for the proposed car-following model with the parameters from Table 1.**

## 4. FUNDAMENTAL DIAGRAM AND EQUILIBRIUM CONDITIONS

The fundamental diagram, i.e., the steady-state relation for speed or flow as a function of density (or spatial gap), is given by the full model with following equilibrium conditions:

- The speeds of all vehicles are the same, and constant over time: i.e.,
  - acceleration $\dot{v}_0 = 0$,
  - speed difference (approaching rate to the leader) $\Delta v = 0$,
- nostochasticity is allowed, i.e., $\beta \to \infty$.

### 4.1. Microscopic Relations

The equilibrium relation is formulated as a relation between the gap $s$ and the speed $v$, e.g., $v = v_e(s)$ or $s = s_e(v)$. Assuming $\frac{\partial \dot{v}}{\partial v} \le 0$ (which should be satisfied for all sensible micro-models) (Treiber and Kesting 2013), the above equilibrium condition leads to:

$$v_e(s) = \min(v_0, v(s)), \qquad \textit{Equation 7}$$

where the steady-state speed $v(s)$ in the interacting range is defined by

$$\dot{v}_{\text{int}}(s, v, \Delta v = 0, \beta \to \infty) = 0. \qquad \textit{Equation 8}$$

**Table1:  Car-Following  Model  Parameters  and  Corresponding  Symbols  for  the Simulation Exercise: the parameters in the top part (above the horizontal line) are the actual model parameters to be calibrated; the parameters in the lower part are secondary parametersthat are not subject to calibration.**

| Parameter | Symbols and Initial Values |
|---|---|
| Sensitivity Exponents of the Generalized Utility | $\gamma = 0.3$ |
| Asymmetry Factor for Negative Utilities | $w_m = 4$ |
| Speed Uncertainty Variation Coefficient | $\alpha = 0.08$ |
| Weighing Factor for Accidents | $w_c = 100000$ |
| Maximum Anticipation Time Horizon | $\tau = 5s$ |
| Logit Uncertainty Parameter (Intra-Driver Variability) | $\beta = 5$ |
| Correlation Time of Intra-Driver Variability | $\tau_{corr} = 20s$ |
| Maximum Acceleration | $a_{max} = 1.5 \text{ m/s}^2$ |
| Desired Speed | $v_0 = 30 \text{ m/s}$ |
| Minimum Gap | $s_0 = 3 \text{ m}$ |
| Acceleration Range Considered Near Interaction Point | $a_0 = 1 \text{ m/s}^2$ |

For the deterministic limit $\beta \to \infty$ ,the interaction acceleration reads

$$\dot{v}_{int} = \arg(\max(U(a;s,v,0))), \qquad \textit{Equation 9}$$

so the steady-state relation can be expressed by

$$U'(a;s,v,0)\,|_{a=0} = 0, \qquad \textit{Equation 10}$$

where the generalized utility $U(a;s,v,\Delta v)$ is understood as a function of the acceleration $a$ .

Setting the seriousness term of the crash utility $\sigma(v,\Delta v) = 1$, one obtains from Equation 3 the condition:

$$U'_{PT}(0) + U'_{crash}(0) = U'_{PT}(0) + w_c\, p'_c(0;s,v,0) = 0 . \qquad \textit{Equation 11}$$

The gradient of the PT utility at $a = 0$ is obtained according to:

$$U'_{PT}(0) = \frac{\left(\dfrac{1+w_m}{2}\right)}{a_0}. \quad \textit{Equation 12}$$

As stated in Section 3 (see Hamdar et al., 2008), the gradient of the utility due to the crash risk is given by:

$$U'_{crash}(a) = -w_c f_N \left( \frac{\Delta v + \frac{1}{2} a \tau - \frac{s - s_0}{\tau}}{\alpha v} \right) \frac{\tau}{2\alpha v}, \qquad \text{Equation 13}$$

or, after inserting the steady-state conditions $a = 0$ and $\Delta v = 0$:

$$U'_{crash}(0) = -\frac{w_c \tau}{2\alpha v} f_N \left( \frac{-(s - s_0)}{\alpha v \tau} \right) = -\frac{w_c \tau}{2\sqrt{2\pi}\alpha v} e^{-\left( \frac{s - s_0}{\sqrt{2}\alpha v \tau} \right)^2}. \qquad \text{Equation 14}$$

Inserting Equations (14) and (13) into (11) and solving for the steady-state gap $s$ results in:

$$s(v) = s_0 + \sqrt{2} v \alpha \tau \sqrt{\ln\left( \frac{a_0 \tau}{v} \right) + \ln\left( \frac{w_c}{2\sqrt{2\pi}\alpha \left( \frac{1 + w_m}{2} \right)} \right)}. \qquad \text{Equation 15}$$

## 4.2. Macroscopic Relations

The congested (interacting) branch of the macroscopic fundamental diagram is obtained by applying the usual micro-macro relations:

$$\rho(v) = \frac{1}{l_{veh} + s(v)}, \qquad \text{Equation 16}$$

and

$$Q(v) = v * \rho(v). \qquad \text{Equation 17}$$

It should be noted that based on the initial parameter testing, the first log term in the square root term of Equation 15 is of the order of unity (unless the speed is very small), while the secondconstant log term is of the order of ten ($w_c$ is in the range of 100 000).

Therefore the equilibrium time headway $T_e(v) = \frac{s(v)}{v}$ is nearly constant (Figure 2) resulting in an approximately triangular fundamental diagram (Figure 3).

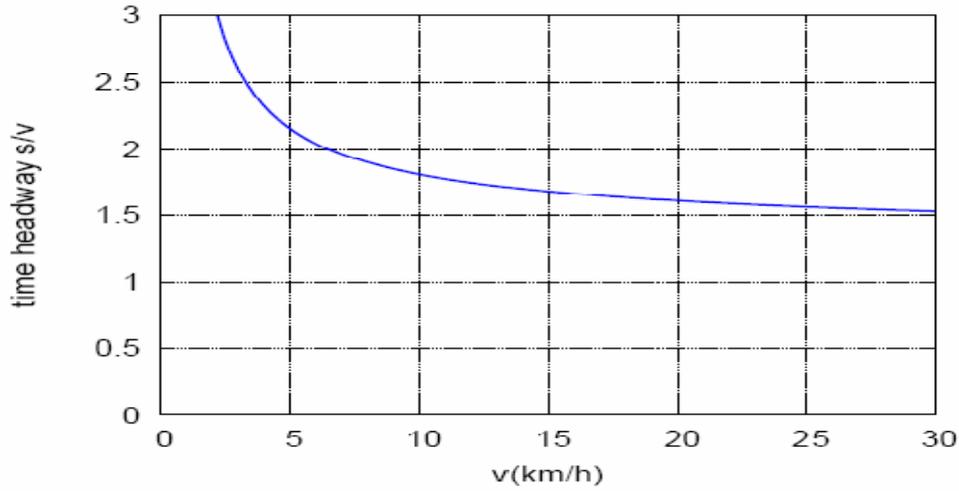

**Figure 2: Equilibrium time headway** $T_e(v) = s(v)/v$ **for the proposed car-following model with the parameters from Table 1.**

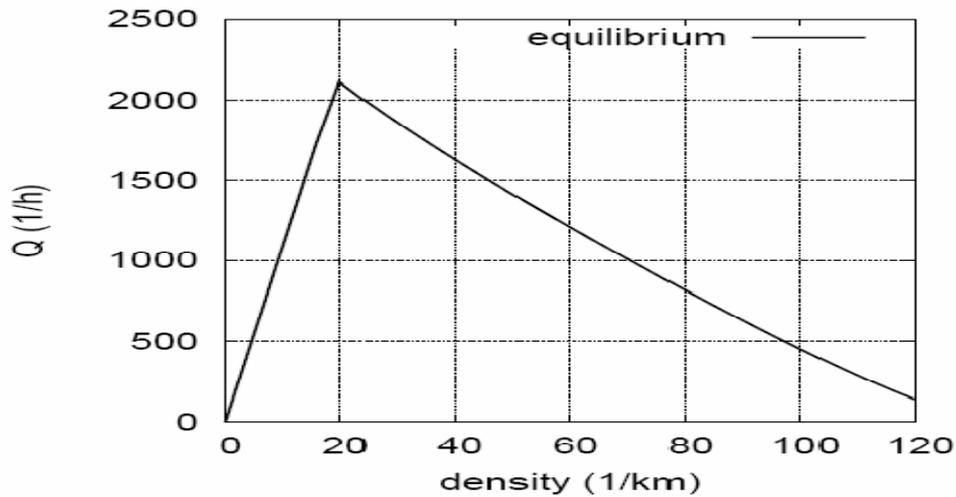

**Figure 3: Fundamental diagram for the accelerationmodel with the parameters from Table 1.**

As in other triangular-shaped fundamental diagrams, the capacity is mainly determined by the inverse of the time headway. The only parameter in the logarithmic term of Equation 15 influencing $s(v)$ (and thus the capacity), and where a variation by several orders of magnitude is probable, is the crash weighing factor $w_c$: all other logarithmic expressions are essentially zero compared to $\ln(w_c)$. Therefore, the effective time headway can be approximated by

$$T_{eff} := \frac{s_e - s_0}{v} \approx \alpha \tau_{max} \sqrt{2\ln(w_c)} \ . \qquad \textit{Equation 18}$$

Figure 4 plots the exact analytic value for $T_{eff}$ calculated based on Equation 15 together with the approximation obtained from Equation 18– both as a function of the speed. One sees that the approximation breaks down only for very low values of $v$. By varying the model parameters while keeping the product defined by Equation 18 constant, once can change the dynamical model properties (e.g., string stability, sensitivities, and accelerations) independently of the static properties which are essentially defined by the capacity.

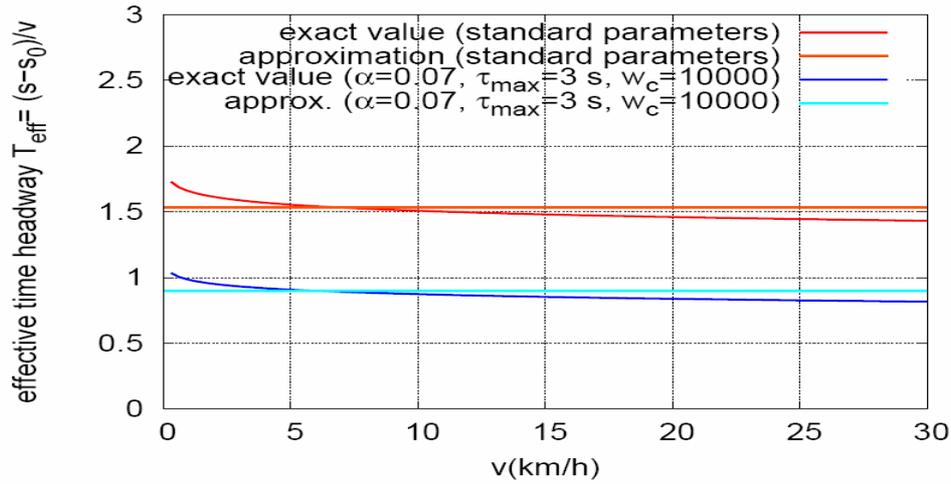

**Figure 4: Effective time headway $T_{eff} = (s_e(v) - s_0)/v$ as a function of the speed in the congested regime. Compared is the exact analytic expression based on Equation 15 with the constant approximation of Equation 18.**

After studying the suggested car-following model in terms of derivation and micro/macro properties, a detailed numerical analysis is presented in the following section.

## 5. NUMERICAL RESULTS: Calibration and Data Analysis

The authors recognize the complexity involved when calibrating a car-following model using trajectory data, if the model's acceleration function is not given explicitly. Multiple alternatives exist when choosing the optimization algorithm, the measures of performance, and the goodness of fit function (Punzo et al., 2012). Even though calibrating the model does not constitute the focus of this study, this section presents a thorough calibration exercise highlighting the car-following model properties.

### 5.1. Data Description and Calibration

To calibrate the model, we have used trajectory data of the Federal Highway Administration's (FHWA) Next Generation Simulation (NGSIM) project (FHWA, 2005 a, b and c); Data for 5678 vehicles have been collected on the 13[th] of April, 2005, on a segment of the Interstate I-80 in Emeryville, San Francisco, USA. The vehicles considered were traveling North-Bound and were tracked using video cameras mounted on the Pacific Park Plaza (a 30 story building located on 6363 Christine Avenue).

The videos were recorded using 7 video cameras (cameras 1 through 7). Camera 1 recorded the southernmost section of the I-80 segment included in the study area and Camera 7 recorded the northernmost section.

Finally, the filtered trajectory points were grouped inthree 15 minutes' intervals:

1- Data Set 1: collected from 4:00 PM to 4:15 PM (2052 vehicles)
2- Data Set 2: collected from 5:00 PM to 5:15 PM (1836 vehicles)
3- Data Set 3: Collected from 5:15 PM to 5:30 PM (1790 vehicles)

Data were recorded every 1/10 second. The area covered in these data sets includes an on-ramp but does not include an off-ramp and has a length of 1650 feet. The focus of this paper is on Data Sets 1 and 2 (FHWA, 2005a and b).

## 5.2. Model Calibration with Heterogeneity

Since the model relies on utility maximization technique with a stochastic choice between different acceleration alternatives, the corresponding equations contain stochastic elements themselves and therefore are analytically intractable.Moreover, this also implies that the objective function (the sum of squared errors) is not smooth as a function of the parameters. For that reason, we have calibrated the model using a nonlinear optimization procedure that is based on a genetic algorithm (Hamdar et al., 2009). The objective is to minimize the deviations between the observedand simulated trajectories when following the same designated leader, and avoiding secondary minima.

Based on the stimuli considered in the acceleration model, the required trajectory data should include speeds for both the leading and the following vehicles of interest. Accordingly, a direct comparison between the measured driver behavior and the trajectories simulated by the car-following model - with the leading vehicle serving as externally controlled input - is possible. In the simulation set-up, the calibration is performed by taking different leader-follower pairs and comparing their driving dynamics with the behavior obtained from the simulated car-following model. The simulated relative speed and the distance gap are initialized to the empirically given relative speed and gap (prescribed values: Treiber and Kesting, 2013):

$$\Delta v^{sim}(t=0) = \Delta v^{data}(0), \qquad\qquad \textit{Equation 19}$$

$$s^{sim}(t=0) = s^{data}(0). \qquad\qquad \textit{Equation 20}$$

The microscopic acceleration model is used to compute the acceleration and thus the trajectory of the following vehicle. The gap to the leading vehicle is computed as the difference between the simulated trajectory $s^{sim}(t)$ (front bumper position) and the recorded position of the rear-bumper of the leading vehicle $s_{lead}^{data}(t)$:

$$s^{sim}(t) = x_{leader}^{data}(t) - x^{sim}(t). \qquad\qquad \textit{Equation 21}$$

The above measure can be directly compared to the gap $s^{data}(t)$ provided by the data. It should be noted that the rear-end and the front-end bumper positions of the leaders and the followers can be extracted since the NGDIM data contains the corresponding vehicle lengths.

In the calibration process, the difference between the observed driving behavior and the driving behavior obtained by the simulated car-following model should be minimized by choosing a set of "optimal" model parameters. Different error measures based on speed, relative speed, or the space gap, can be used. Normally, the error in the space gap $s$ is adopted: when optimizing with respect to $s$, the average relative speed errors are automatically reduced. In contrast, when optimizing with respect to the relative speeds $\Delta v$, the error in the distance gap may incrementally grow (Punzo et al., 2012).

In this study, due to errors in recording the space gaps in the NGSIM data (Thiemann et al., 2008), the optimization procedure is performed with respect to the speed $v$; since the NGSIM data are collected during the peak-hour PM congested period, the image processing of the recorded videos resulted in transforming some space headways into negative space gaps after subtracting the vehicle lengths. On the other hand, since the form of the objective function has a direct impact on the calibration results, three different error measures can be considered. The relative error is defined as a function of the empirical and the simulated time series ($v^{sim}(t)$ and $v^{data}(t)$ respectively):

$$F_{rel}\left[v^{sim}\right] = \sqrt{\left\langle\left(\frac{v^{sim} - v^{data}}{v^{data}}\right)^2\right\rangle}, \qquad \text{Equation 22}$$

where $\langle.\rangle$ refers to the temporal average of a time series of duration $\Delta T$:

$$\langle z\rangle = \frac{1}{\Delta T}\int_0^{\Delta T} z(t)dt. \qquad \text{Equation 23}$$

The relative error is more sensitive to small speeds $v$ than to large speeds. The main reason behind such a sensitivity is that the measure is weighted by the inverse recorded speed $v^{data}$.

The second measure considered is the absolute error:

$$F_{abs}\left[v^{sim}\right] = \sqrt{\frac{\left\langle\left(v^{sim} - v^{data}\right)^2\right\rangle}{\left\langle v^{data}\right\rangle^2}}. \qquad \text{Equation 24}$$

Since the denominator is averaged over the whole trajectory interval, the absolute error $F_{abs}\left[v^{sim}\right]$ is less sensitive to small deviations from the empirical data than the relative error $F_{rel}\left[v^{sim}\right]$. On the other hand, the absolute error is more sensitive to large differences in the numerator (large speeds reflecting large gaps). It should be noted that the error measures are normalized so they are independent from the duration $\Delta T$.

Since the absolute error systematically overestimates errors for large gaps (at high speeds) while the relative error systematically overestimates deviations of the observed headway in the low speed range, a mixed error measure will be used as the objective function in this paper:

$$F_{mix}\left[v^{sim}\right] = \sqrt{\frac{1}{\left|v^{data}\right|}\left\langle\frac{\left(v^{sim}-v^{data}\right)^2}{\left|v^{data}\right|}\right\rangle}. \qquad \textit{Equation 25}$$

Once the objective function to be minimized is defined, the genetic algorithm is applied as a search heuristic to find an approximate solution to the nonlinear optimization problem:

   i-   A "chromosome" represents a parameter set of the car-following model introduced earlier in this work and a population consists of $N_{GA}$ such chromosomes.
   ii-   In each chromosome generation, the fitness of each chromosome is determined via the objective function defined in Equation 25.
   iii-   All pairs of chromosomes are exclusively generated from the current population and recombined to generate new chromosomes.
   iv-   The cross-over point where two chromosomes are combined is randomly selected.
   v-   Except for the chromosome with the best fitness score, all the genes (model parameters) are mutated (varied randomly) following a given probability. The resulting chromosomes (new generation) are used in the next iteration.
   vi-   Initially, a fixed number of generations is evaluated. The evolution is then terminated when the best-of-generation score converges from one iteration to another for a given number of generations.

In this research, the initial set of parents (10 parents) is initiated where the parameters are given values in the proximity of those provided and tested in Table 1.At each iteration, these parents produce 90 children chromosomes where the best 10 candidates of the $N_{GA}$ population ($N_{GA} = 90 + 10$) are kept to the next iteration. The calibration process continues until no improvement of more than 0.01 is observed for 20 consecutive iterations, or when the error reaches the threshold of 15%. It should be noted that a mutation rate of 10% is applied in all iterations.

As seen in Table 1, there are seven main parameters to be calibrated. The additional parameter incorporated in the process is the reaction time. Focusing on the car-following instances in the offered trajectory data,the related summary results of the calibration exercise using data sets 1 and 2 are shown in Tables2 and3(i.e; all vehicle trajectories). The presented parameters arethe calibrated parameters that led toerrors below a 30%threshold. For illustration reasons, sample simulated versus trajectory data are shown in Figure 5.Notice that towards the end of the calibration simulation (aroundtime-step 1000 in Figure 5), the simulated speeds and the simulated space gaps drop to zero. The time of this drop corresponds to the time when gap data ceases to exist in the NGSIM data sets(i.e., the cameras cannot detect anymore the corresponding lead vehicle). When such lead-vehicle data are not available (space gaps and relative speeds), the calibration exercise is terminated and no further contribution to the mixed error term is recorded.

The first interesting finding is the important level of inter-driver heterogeneity although the distributions of parameters values are not clear Gaussian distributions. A clear peak appears in all distributions but with a vast range of parameters values. When examining the average values, the cognitive nature of parameters allows interesting interpretations; for example, drivers seem to put 4 times the negative weight (Wm ~ 4) on losses in speed than on gains (the corresponding weight is assumed to be equal to 1). Moreover, even though crashes are not avoided through the use of safety constraints, the calibrated high value of the crash weight Wc (~100000) reduces the possibility of accidents in this simulation exercise. On the other hand, notice that the mean and the standard deviation for the *Gamma* and the reaction time*Rt* parameters are close in value. This may suggest a possible exponential probability density function. Also notice that the correlation matrix shows mainly low correlation values between different parameters. This may indicate a high level of independence between parameters which is a desirable property. The corresponding parametric correlation is a subject of future research.

The distributions of the different calibrated parameters values are illustrated in Figures 6 (Data Set 1) and 7(Data Set 2); one may point that the *Gamma* parameter and the *Rt* parameter may have exponential or Weibull distribution functions(equal mean and standard deviation) while the rest of the parameters havea non-Gaussian shaped distribution functions with a governing peak value.

**Table 2: Summary Statistics for Calibrated Parameter Values Using Genetic Algorithm (GA) – Data Set 1.**

| Parameter | Units | Mean | Std. Dev. | Minimum | Maximum |
|---|---|---|---|---|---|
| *Gamma* ($\gamma$) | - | 0.333991 | 0.339046 | 0 | 1.9 |
| *Wm* ($w_m$) | - | 3.97208 | 2.6452 | 0.2 | 9.8 |
| *Wc* ($w_c$) | - | 97077.4 | 21143.6 | 50000 | 149000 |
| *T* ($\tau$) | seconds | 5.08938 | 1.98571 | 1 | 10.9 |
| *Alpha* ($\alpha$) | - | 7.74E-02 | 3.93E-02 | 1.00E-02 | 0.46 |
| *Beta* ($\beta$) | - | 5.32671 | 2.097 | 1 | 10.9 |
| *Tcorr* ($\tau_{corr}$) | seconds | 19.9833 | 4.53906 | 10 | 29 |
| Reaction Time (*Rt*) | seconds | 0.587102 | 0.688398 | 0.1 | 3.2 |

**Table 3: Correlation Matrix for Calibrated Parameters Using GA– Data Set 1.**

| Correlation | Gamma | Wm | Wc | Tmax | Alpha | Beta | Tcorr | Rt |
|---|---|---|---|---|---|---|---|---|
| Gamma | 1 | 0.26 | -0.07 | 0.4 | -0.14 | 0.2 | 0.01 | -0.009 |
| Wm | 0.26 | 1 | -0.02 | 0.15 | -0.09 | 0.11 | 0.05 | -0.04 |
| Wc | -0.07 | -0.02 | 1 | -0.07 | 0.001 | -0.03 | 0.01 | -0.002 |
| T | 0.4 | 0.15 | -0.07 | 1 | -0.18 | 0.04 | -0.06 | 0.19 |
| Alpha | -0.14 | -0.09 | 0.001 | -0.18 | 1 | -0.08 | -0.00007 | -0.02 |
| Beta | 0.2 | 0.11 | -0.03 | 0.04 | -0.08 | 1 | -0.03 | 0.07 |
| Tcorr | 0.01 | 0.05 | 0.01 | -0.06 | -0.00007 | -0.03 | 1 | 0.04 |
| Rt | -0.009 | -0.04 | -0.002 | 0.19 | -0.02 | 0.07 | 0.04 | 1 |

**Table 4: Summary Statistics for Calibrated Parameter Values Using Genetic Algorithm (GA) – Data Set 2.**

| Parameter | Units | Mean | Std. Dev. | Minimum | Maximum |
|---|---|---|---|---|---|
| Gamma ($\gamma$) | - | 0.31619 | 0.345064 | 0 | 1.9 |
| Wm ($w_m$) | - | 3.85276 | 2.62247 | 0.2 | 9.9 |
| Wc ($w_c$) | - | 97556.2 | 22059.5 | 50000 | 149000 |
| T ($\tau$) | seconds | 5.04476 | 1.95507 | 1 | 10.9 |
| Alpha ($\alpha$) | - | 7.72E-02 | 3.54E-02 | 1.00E-02 | 0.45 |
| Beta ($\beta$) | - | 5.37314 | 2.33261 | 1 | 10.8 |
| Tcorr ($\tau_{corr}$) | seconds | 20 | 4.45203 | 10 | 29 |
| Reaction Time (Rt) | seconds | 0.658857 | 0.726583 | 0 | 2.9 |

**Table 5: Correlation Matrix for Calibrated Parameters Using GA– Data Set 2.**

| Correlation | Gamma | Wm | Wc | Tmax | Alpha | Beta | Tcorr | Rt |
|---|---|---|---|---|---|---|---|---|
| Gamma | 1 | 0.23 | 0.04 | 0.34 | -0.11 | 0.16 | -0.03 | -0.02 |
| Wm | 0.23 | 1 | 0 | 0.19 | -0.11 | 0.13 | 0.03 | -0.05 |
| Wc | 0.04 | 0 | 1 | 0.02 | 0.03 | 0.02 | -0.07 | 0.06 |
| T | 0.34 | 0.19 | 0.02 | 1 | -0.24 | 0.13 | 0.05 | 0.16 |
| Alpha | -0.11 | -0.11 | 0.03 | -0.24 | 1 | -0.07 | -0.06 | 0.02 |
| Beta | 0.16 | 0.13 | 0.02 | 0.13 | -0.07 | 1 | 0.00 | 0.12 |
| Tcorr | -0.03 | 0.03 | -0.07 | 0.05 | -0.06 | 0.00 | 1 | 0 |
| Rt | -0.02 | -0.05 | 0.06 | 0.16 | 0.02 | 0.12 | 0 | 1 |

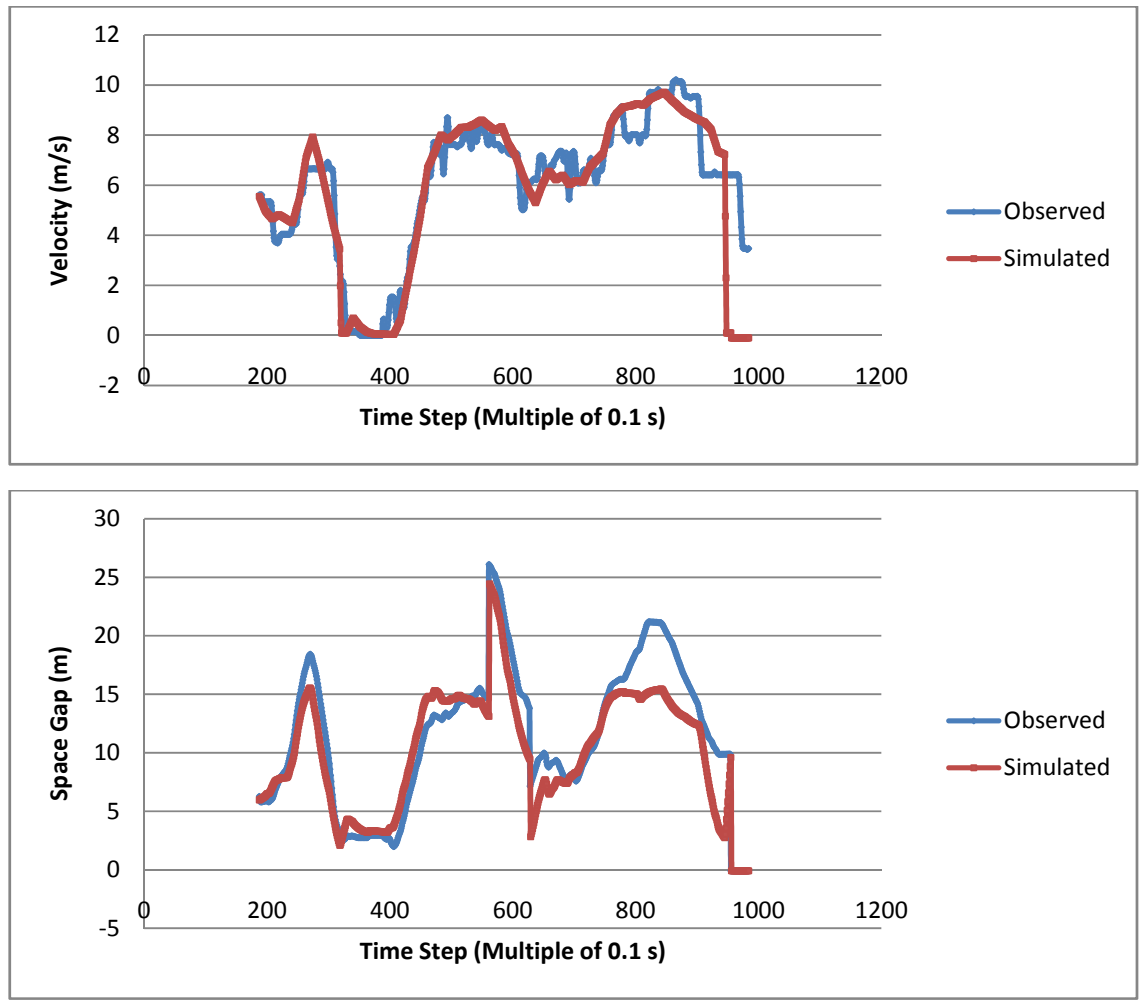

**Figure 5: Simulated versus observed speeds (upper graph – mixed error = 0.12) and space gaps (lower graph – mixed error = 0.29) for Vehicle 32 from data set 1**

However, when trying to estimate the corresponding distributions, no statistically significant function was found. In other words, at this stage, due to the lack of data, conclusive results on the distribution followed by each parameter values could not be reached (null hypothesis on the corresponding distributions tested for acceptance or rejection); In addition to the lack of data, themain reason behind such result is that, even though a significant heterogeneity (spread of parameter values across drivers) exists, the concentration of parameter values around one peak is too high for existing parametric distribution functions to capture.

**5.3. Model Validation**

In this section, we perform a simple validation to assess the robustness of the parameter values calibrated in the previous section. Since the acceleration model parameters (Table 1) are driver specific, and data sets 1 and 2 have a different number of vehicles, the parameter values corresponding to the peak values found in the calibration process are applied in the validation process. The validation results are provided in Table 7.

**Table 7: Validation Error Terms**

| Validation Error | Mean Mixed Error | Std of Mixed Error | Minimum | Maximum |
|---|---|---|---|---|
| Data Set 1 | 0.3258 | 0.353026 | 0.036213 | 1.965293 |
| Data Set 2 | 0.3107 | 0.291286 | 0.048659 | 2.191862 |

For both data sets 1 and 2, the mean mixed error term is equal to ~0.3. Compared to the error threshold of 15% specified in the genetic algorithm procedure, the validation error term is almost double the mean error found in the calibration exercise. This result indicates the significance of inter-driver heterogeneity.

Some of the validation errors reach values close to 200%. To examine this phenomenon, the distribution of the error term is plotted for data sets 1 and 2 in Figure 8; around 50% of the errors have a value less than 30% (~50% of the error values less than 30%). This error threshold is comparable to that found in existing calibration studies (Kesting and Treiber, 2008) and is considered reasonable. For the remaining 50%, the main problem is the deterministic nature of assigning calibrated parameter values (peak values) to different drivers irrespective of their behavioral nature found in the calibration process (their initial calibrated parameters).In addition, inter-driver variations pertain to the discussed distributions of the parameters; an upper limit of the intra-driver variations can be assessed by the residual sum of squared errors (SSE). In terms of variances and assuming independence between inter- and intra-drive variations, the cross-calibration/validation variance (or $SSE_{cross}$) can be used to estimate the relative contributions:

$$SSE_{cross} = SSE_{intra} + SSE_{inter} \qquad \textit{Equation 26}$$

Where $SSE_{intra}$ (the average SSE of all trajectories on calibration) characterizes the upper limit of the intra-driver variations, and $SSE_{inter}$ represents inter-driver variations. After filtering the trajectories that led to determinate error values, $SSE_{cross}$ and $SSE_{intra}$ were computed for data sets 1 and 2 and the following was found:

Data Set 1: $SSE_{cross}$ = 548.7836 and $SSE_{intra}$ = 118.3996

Data Set 2: $SSE_{cross}$ = 390.0792 and $SSE_{intra}$ = 102.7188

Such results indicate the considerable contribution of inter-driver heterogeneity to the cross-validation error variance.

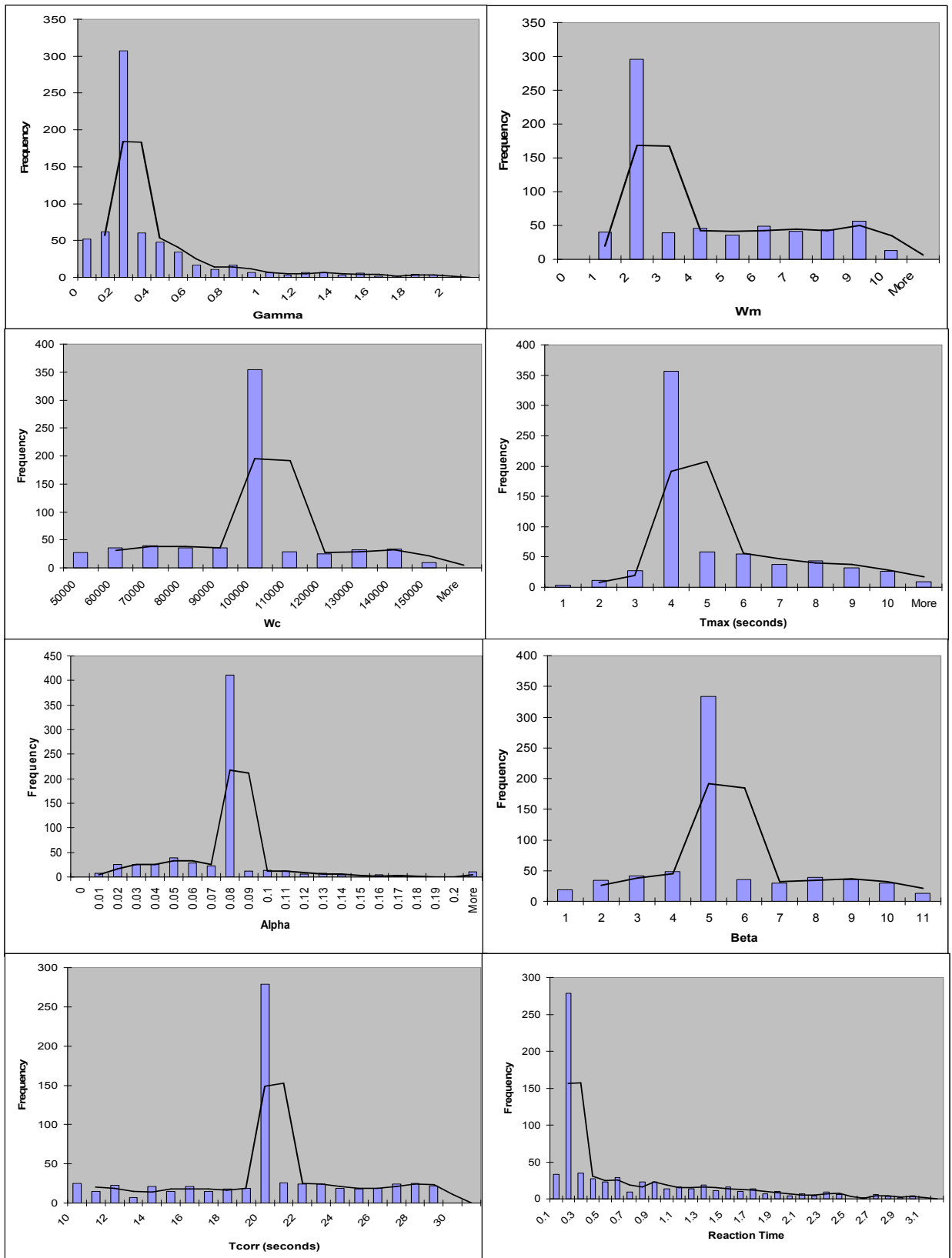

**Figure 6: Distribution of Utility-Based Model Parameters Using GA – Data Set 1**

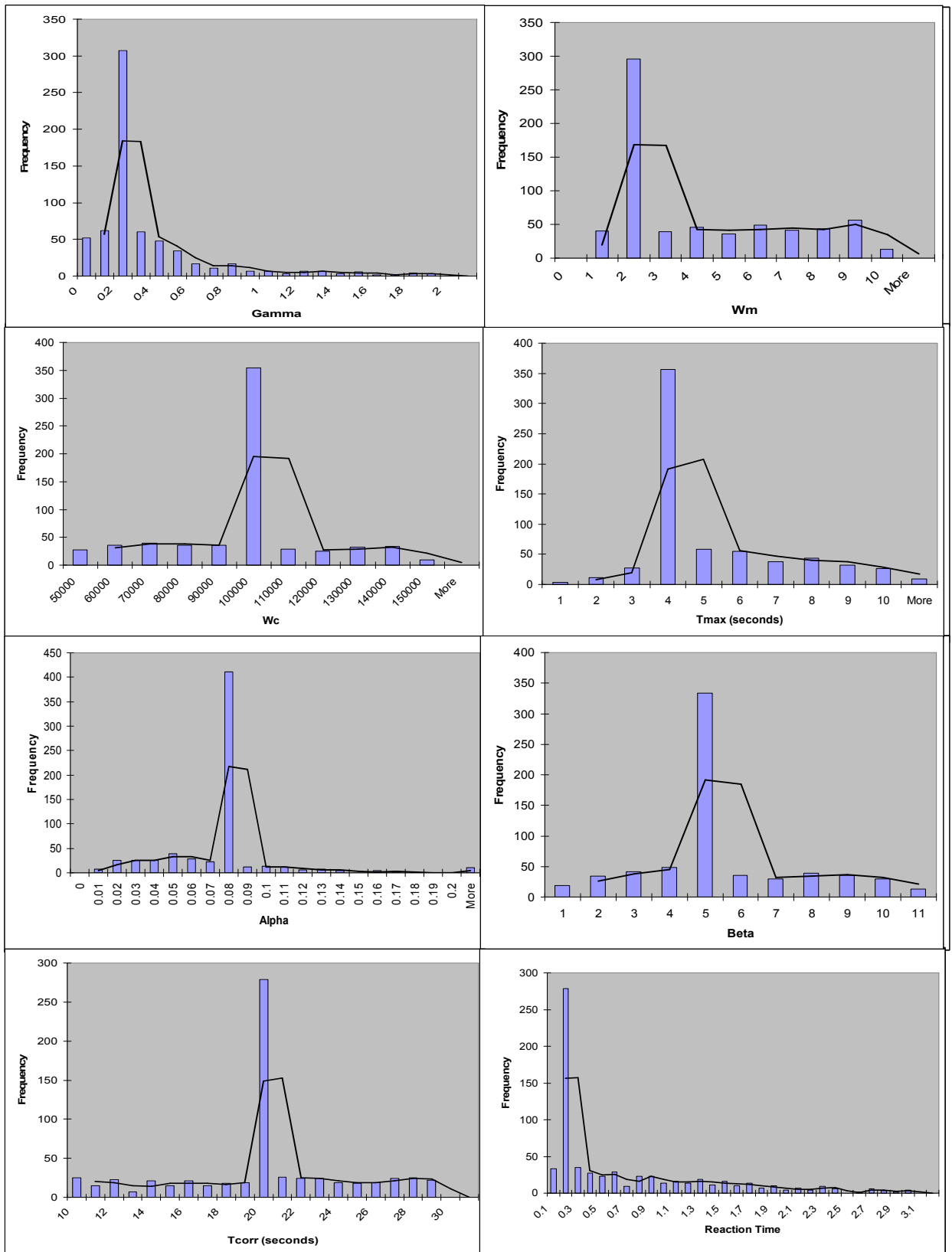

**Figure 7:Distribution of Utility-Based Model Parameters Using GA – Data Set 2**

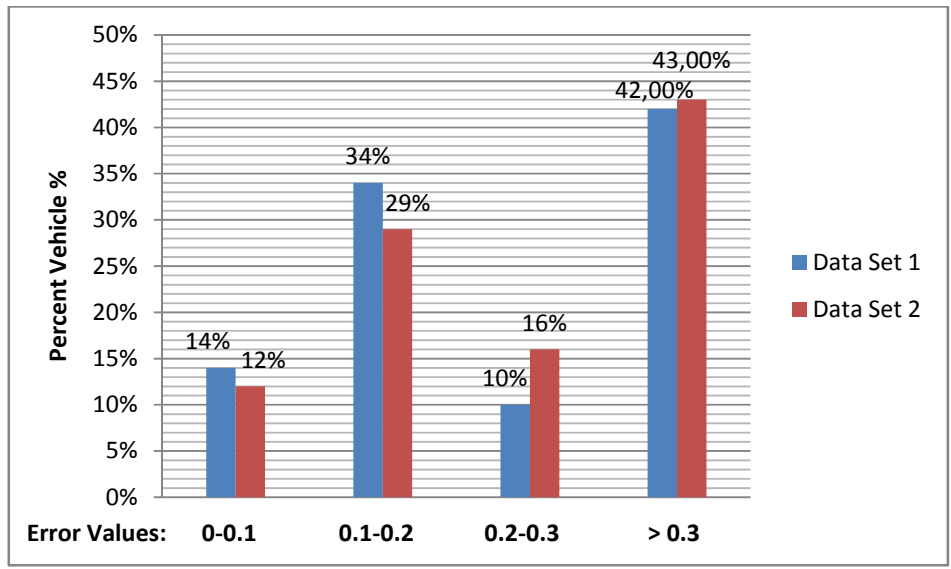

**Figure 8: error distribution across vehicles when using parameters of data set 2 on data set 1 (Data Set 1), and using parameters of data set 1 on data set 2 (data set 2).**

## 6. SIMULRATION AND SENSITIVITY ANALYSIS

### 6.1. Flow-Density Relation

In this section, the car-following model is simulated using parameters values of the same order of the estimates found in the previous section. The car-following model is combined in this exercise with the MOBIL lane changing model for added robustness (Kesting et al., 2007): the acceleration rules are used for assessing the "safe" comfortable gaps to change lanes. The vehicles are "injected" on a two-lane freeway and a one-lane on-ramp merging together at location x = 10 km. The "freeway entrance" is at x=0 km, and x=6 km, 9 km, and 10 km are the positions of virtual detectors. The calibrated lane-changing function is called at each simulation time step to determine the desirability of changing lanes. Figure 9illustrates the resulting fundamental diagram.

In the fundamental diagram, analytically, two main regions emerge: a free-flow region (green straight line) and a congested region (red straight line). When simulating the model, virtual one-minute detectors are placed to collect flow and density measures in three different scenarios based on the location of the merging of the on-ramp traffic and the main-stream traffic. When this location is close to the "freeway entrance" (x = 6 km), the transition between the free-flow region and the congested region is seen through a sharp traffic breakdown (sudden drop in volume – red line). This traffic breakdown is followed by scattered "non-synchronized" flow-density points in the congested region. As the on-ramp is further away from the free-way entrance, a smoother transition occurs where synchronized flow-density points appear. This kind of traffic dynamics imitate some of the observations that appear in real-life situations (Treiber et al., 2000).

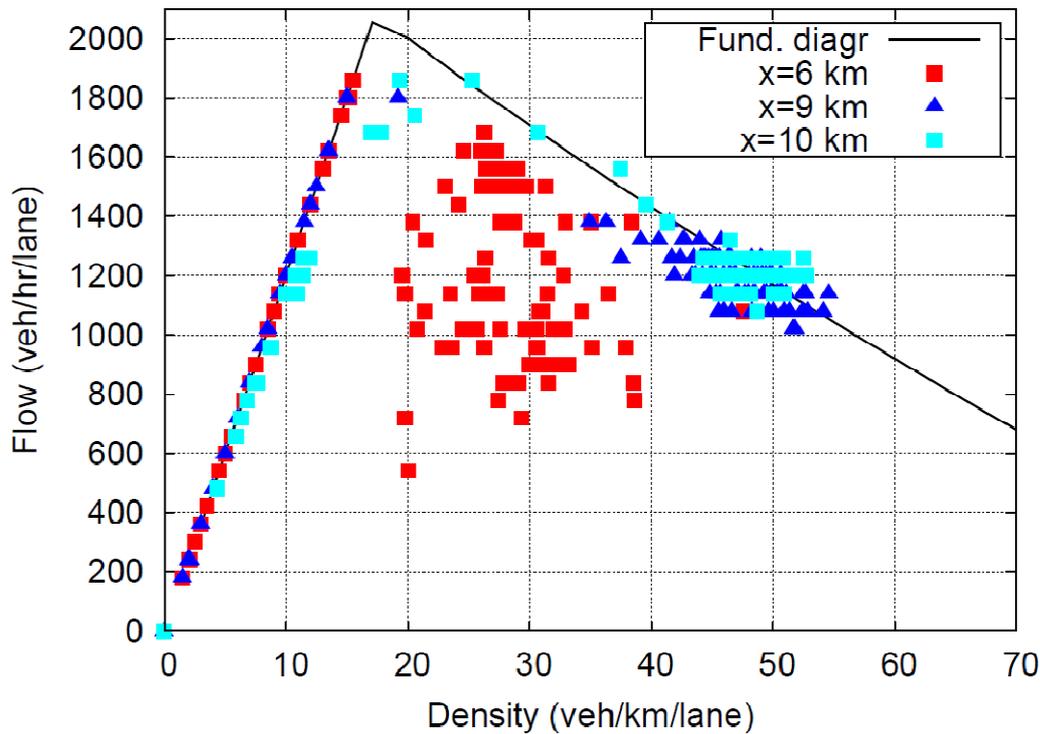

**FIGURE 9.Fundamental diagram of the combined car-following/MOBIL model with the parameters described in Table 1. The results are based on virtual one-minute detectors at the indicated locations. The on-ramp bottleneck is at x=10 km.**

### 6.2. Inter-Driver Heterogeneity

In the inter-driver heterogeneity related simulation sensitivity analysis, the base-case scenario is taken with the calibrated parameters. The main cognitive parameters of interest are the crash-penalty, the negative coefficient associated with losses in speed, the positive coefficient associated with gains in speed, the driver's uncertainty, the anticipation time and the reaction time.

The acceleration model is simulated with a simulation time-step of 0.1 second .The vehicles are "injected" into a 10 km two-lane freeway section. The initial flow-rate is controlled by an exponential inter-arrival time with a given mean. Since the interest is in capturing all the regions of the fundamental diagram (free-flow and uncongested), at different road sections, a kilometer bottleneck is created through the allowance of an unstable and abrupt vehicle deceleration. This also favor the creation of rear-end collision for testing the influence of the different model parameters. Figure 10-a shows the resulting flow-density data points and hysteresis triangle if formed accordingly.

To test the effect of inter-driver heterogeneity, two families of scenarios are offered. The first family is related to homogenous traffic where the parameters values for all vehicles are constant and correspond to the peak of values of the parameters distributions found in the calibration exercise. The second family is related to heterogeneous traffic where the parameters values have a normal distribution where the mean corresponds to the peak found in the calibration results; Figure 10-b shows a slightly increasing flow-density data points scattering if compared to the homogeneous scenario simulation.Such slight increase in scattering of the flow-density data points while keeping the corresponding triangular fundamental diagram characteristics underlines the robustness of the model.

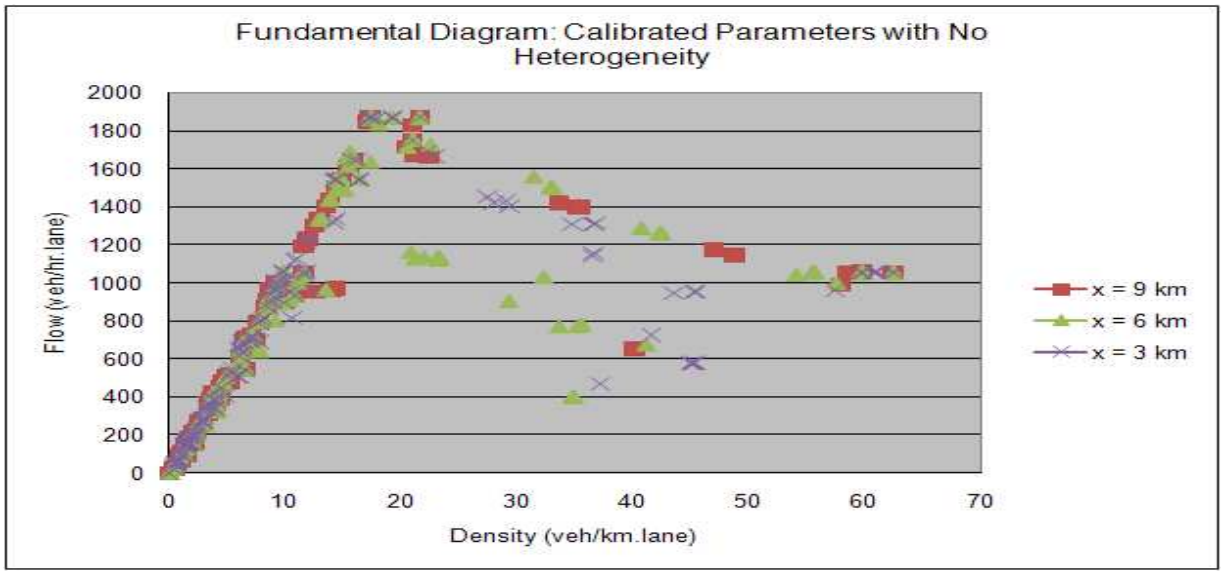

**Figure10-a: Flow-density relation.The results are based on virtual one-minute detectors at the indicated locations.**

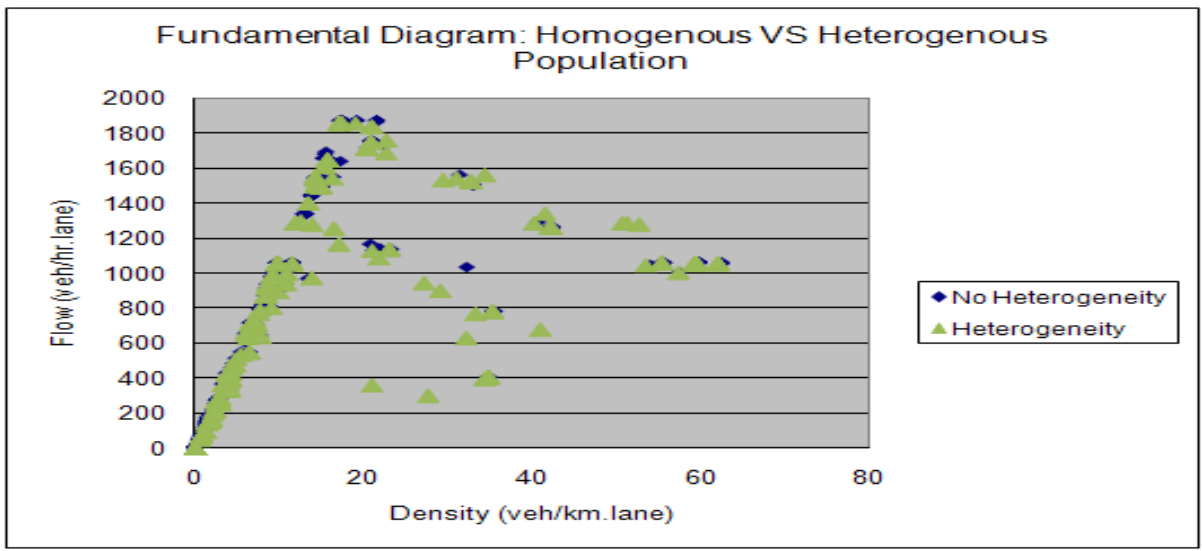

**Figure10-b: Flow-density relation with and without heterogeneity**

**6.3. Crash-Investigation**

Further extensive sensitivity analysis is performed to test the effect the parameters values and heterogeneity on the crash creation and distribution. The basic results are shown in Table 8.

Incidents are created while there is an interplay between the weight parameters (Wm, W+ and Wc) and the time parameters (Tau = anticipation time and RT = reaction time). Regarding the weight parameters, the relative decrease of Wm (a lesser value than 0.098 < 1) with respect to W+ (= 1 initially) starts producing incidents. On the other hand, when increasing W+ to 2 (until we have a linear value function with Wm = W+), no incidents are created even when tailgating is favored and higher throughput is observed: the incident creation in rear-end collisions is related to a deceleration behavior rather than an acceleration behavior. Finally, as observed in the homogeneous case (2nd row), the relative weight between Wm and W+ contribute in creating the incident while the crash weight's (Wc) role seems to be producing the flow dynamics and instability; such instabilities (stop and go, tailgating) constitutes an encouraging environment for incident scenarios. On the other hand, reaction time and anticipation time contributes for the creation of incident in a different manner; when comparing the reaction time scenarios in the heterogeneous traffic versus the homogeneous traffic, it is the heterogeneity in the reaction time that produced the highest number of incidents. This may suggest the role of correlation between two successive vehicles in stabilizing traffic conditions. As for the anticipation time, when low anticipation times are used or when a high discrepancy between the anticipation times of two successive vehicles exists, the probability of incidents creation increase.Accordingly, inter-driver heterogeneity is a major aspect than needs to be understood when studying safety in vehicular traffic.

**Table8:  Sensitivity Analysis Results: Heterogeneity VS Crash Distribution.**

| Bottleneck Scenario | Inter-Driver Heterogeneity | | | |
|---|---|---|---|---|
| | Inter-Driver Heterogeneity 1 | | Inter-Driver Heterogeneity 2 | |
| Parameter | (mean, std, range) | # Accidents | (mean, std, range) | # Accidents |
| Wm | 1, 0.5, 2 | 30 | 0.5, 0.5, 1 | 9 |
| W+ | 1, 0.5, 2 | 0 | 2, 0.5, 2 | 0 |
| Wc | 100000, 10000, 40000 | 0 | 100, 50, 200 | 0 |
| Beta Utility Uncertainty | 5, 0.5, 2 | 0 | 100, 25, 100 | |
| Tau | 4, 0.5, 2 (1.3 sec) | 0 | 4, 1.5, 7 (1.3 sec) | 4 |
| Reaction Time | 2, 1, 3.8 | 249 | 1, 0.5, 0.9 | 0 |
| Bottleneck Scenario | Constant Change - All Vehicles | | | |
| | Change 1 | | Change 2 | |
| Parameter | Value | # Accidents | Value | # Accidents |
| Wm | 1 | 0 | 3 | 0 |
| W+ | 2 | 0 | 5 | 0 |
| Wc | 50000 | 0 | 500 | 0 |
| Beta Utility Uncertainty | 100 | 0 | 0.1 | 0 |
| Tau | 2 | 0 | 1 | 0 |
| Reaction Time | 5 | 27 | 1 | 0 |
| | Change 3 | | Change 4 | |
| Wm | 0.1 | 9 | 0.5 | 0 |
| W+ | 20 | 0 | 100 | 0 |
| Wc | 5 | 0 | 0.5 | 0 |
| Beta Utility Uncertainty | 100000 | 0 | 0 | 0 |
| Tau | 1 (inter-arrival 0.1) | 0 | 1  (1.3 seconds  RT) | 4 |
| Reaction Time | 1.3 | 0 | 2 | 0 |

## 7. CONCLUSIONS AND FUTURE RESEARCH NEEDS

In this paper, studying and calibrating a stochastic car-following model while testing it for validity in terms of capturing congestion dynamics and incidents creation is presented. Based on the extensive numerical analysis, this study showed that the GA approach is suitable to calibrate car-following models with complex structures and capturing inter-driver heterogeneity. Using a utility-based stochastic model, the study shows that inter-driver heterogeneity exists with different weights attributed to the gains and losses associated with different acceleration terms ($w_m$ and $w_c$ of Table 1). Such phenomena contributes to a higher scattering of flow-density data points (instabilities) with little influence on the capacity (~1900 vehicles/hr/lane) and the triangular shape of the resulting fundamental diagram (hysteresis triangle).

After testing the model for incidents creation, the presented utility-based structure seems to be more resistant and complex (cognitive aspect) than existing models when two vehicles collide (Hamdar and Mahmassani, 2008). Both individual and chain type accidents can be produced using weight parameters and time parameters. However, incident creation is not based on simple relaxation of safety constraints especially that there are none. Incident creations are based on both parameters values change (interplay of negative weight parameter, positive weight parameter and crash parameter) and parameters values heterogeneity (reaction time) and inter-vehicle parameters values

In future studies, this calibration exercise should be performed on different data sets where the inter-driver dynamics are recorded for longer durations and on a longer stretch of freeway. Calibration should be performed while considering intra-driver heterogeneity and parameters inter-correlation. Finally, the incident-related sensitivity analysis should be generalized to include the fixed object crashes and the results need to be calibrated with real-life incidents scenarios.

## 8. ACKNOWLEDGEMENTS

This material is based upon work supported by the National Science Foundation under Grant No. 0927138. Any opinions, findings, and conclusions or recommendations expressed in this material are those of the author(s) and do not necessarily reflect the views of the National Science Foundation.